\documentclass[conference]{IEEEtran}
\IEEEoverridecommandlockouts
\usepackage{cite}
\usepackage{amsmath,amssymb,amsfonts}
\usepackage{algorithm}
\usepackage{algpseudocode}
\usepackage{graphicx}
\usepackage{textcomp}
\usepackage{xcolor}
\usepackage{hyperref}
\usepackage{multirow}
\usepackage{booktabs}
\usepackage{orcidlink}

\hypersetup{
    colorlinks=true,
    linkcolor=blue,
    filecolor=magenta,      
    urlcolor=blue,
    citecolor=blue
}

\def\BibTeX{{\rm B\kern-.05em{\sc i\kern-.025em b}\kern-.08em
    T\kern-.1667em\lower.7ex\hbox{E}\kern-.125emX}} 

\IEEEoverridecommandlockouts

\begin{document}

\title{Evolving a multi-population evolutionary-QAOA on distributed QPUs}

\makeatletter
\newcommand{\linebreakand}{%
  \end{@IEEEauthorhalign}
  \hfill\mbox{}\par
  \mbox{}\hfill\begin{@IEEEauthorhalign}
}
\makeatother

\author{

\IEEEauthorblockN{1\textsuperscript{st} Francesca Schiavello\,\orcidlink{0009-0003-2651-2856}}
\IEEEauthorblockA{\textit{The Hartree Centre, STFC}
\\Sci-Tech Daresbury, Warrington, UK
\\\href{mailto:francesca.schiavello@cern.ch}{francesca.schiavello@cern.ch}}

\and

\IEEEauthorblockN{2\textsuperscript{nd} Edoardo Altamura\,\orcidlink{0000-0001-6973-1897}}
\IEEEauthorblockA{\textit{The Hartree Centre, STFC}
\\Sci-Tech Daresbury, Warrington, UK
\vspace{5pt}
\\\textit{Yusuf Hamied Department of Chemistry}
\\University of Cambridge, Lensfield Road, 
\\Cambridge CB2 1EW, United Kingdom
\\\href{mailto:edoardo.altamura@stfc.ac.uk}{edoardo.altamura@stfc.ac.uk}}

\and

\IEEEauthorblockN{3\textsuperscript{rd} Ivano Tavernelli\,\orcidlink{0000-0001-5690-1981}}
\IEEEauthorblockA{\textit{IBM Research Zurich} 
\\ 8803 R{\"u}schlikon, Switzerland
\\\href{mailto:ita@zurich.ibm.com}{ita@zurich.ibm.com}}

\linebreakand

\IEEEauthorblockN{4\textsuperscript{th} Stefano Mensa\,\orcidlink{0000-0002-0938-144X}}
\IEEEauthorblockA{\textit{The Hartree Centre, STFC}
\\Sci-Tech Daresbury, Warrington, UK
\\\href{mailto:stefano.mensa@stfc.ac.uk}{stefano.mensa@stfc.ac.uk}}

\and

\IEEEauthorblockN{5\textsuperscript{th} Benjamin Symons\,\orcidlink{0000-0001-5742-1082}}
\IEEEauthorblockA{\textit{The Hartree Centre, STFC}
\\Sci-Tech Daresbury, Warrington, UK
\\\href{mailto:benjamin.symons@stfc.ac.uk}{benjamin.symons@stfc.ac.uk}}

}

\maketitle

\begin{abstract}

Our work integrates an Evolutionary Algorithm (EA) with the Quantum Approximate Optimization Algorithm (QAOA) to optimize ansatz parameters in place of traditional gradient-based methods. We benchmark this Evolutionary-QAOA (E-QAOA) approach on the Max-Cut problem for \textit{d}-3 regular graphs of 4 to 26 nodes, demonstrating equal or higher accuracy and reduced variance compared to COBYLA-based QAOA, especially when using Conditional Value at Risk (CVaR) for fitness evaluations. Additionally, we propose a novel distributed multi-population EA strategy, executing parallel, independent populations on two quantum processing units (QPUs) with classical communication of `elite' solutions. Experiments on quantum simulators and IBM hardware validate the approach. We also discuss potential extensions of our method and outline promising future directions in scalable, distributed quantum optimization on hybrid quantum-classical infrastructures.

\end{abstract}


\section{Introduction}
Quantum Approximate Optimization Algorithms (QAOAs) \cite{Farhi} are promising methods for solving combinatorial optimization problems~\cite{farhi2016quantum, Blekos, Abbas, Symons, hadfield2019quantum} on quantum computers. Some of the first implementations of QAOA \cite{PhysRevA.97.022304, PhysRevX.10.021067} demonstrated the potential for quantum advantage as it outperformed the best-known classical algorithm to solve the Max-Cut problem at the time, assuming the use of optimal parameters \cite{PhysRevA.103.042612}. Since then, classical computing has taken greater strides to maintain an advantage over quantum optimization. Yet, the QAOA continues to be an active area of research for the quantum computing community~\cite{Zhou}. The QAOA is a variational algorithm~\cite{peruzzo2014variational,CerezoVQA} and therefore faces the well-known challenges of barren plateaus and noise degradation~\cite{CerezoBP, Wang} that occur with increasing circuit width and depth in the current era of noisy, near-term quantum computers. These issues make it increasingly hard for optimizers to move in a search landscape as the problem grows to a relevant scale \cite{Guerreschi, Harrigan}.
These difficulties motivate our research into an Evolutionary Algorithm (EA) as an optimizer for QAOA, instead of linear or gradient-based methods. There is still a discussion within the literature about whether or not gradient-free methods are useful, for example, Arrasmith \textit{et al.} argue that these do not solve the barren plateau problem~\cite{Arrasmith}. However, others state that Genetic Algorithms (GAs) and EAs have an advantage for faster convergence~\cite{Ibarrondo}, avoiding barren plateaus~\cite{Nádori}, and beating the state-of-the-art gradient-free optimization for better accuracy~\cite{Acampora}. Note that there is no clear consensus in the field on the distinction between GA and EA and the two have overlapping definitions~\cite{Eiben}. As such, a benefit for one would clearly benefit the other. Furthermore, EAs have been shown to handle complex optimization problems with noisy or uncertain fitness functions~\cite{Eiben,Holland}. If these conditions hold, and EAs prove to be more robust as the literature continues to tackle larger-scale problems, this would make them a natural choice in combining them with QAOAs and other Variational Quantum Algorithms (VQAs).  
  
In addition, EAs offer the potential for embarrassingly parallel applications, one specific method developed is that of Multi-population Genetic Algorithms (MGAs)~\cite{Shi, Belding}, which evolve independent populations in isolation. MGAs have been shown to be successful in avoiding premature convergence, an issue that sometimes arises in GAs, and increasing diversity within the population's evolution, a key factor for successful convergence \cite{Shi, Gong, Shi2}. We present a technique that implements an MGA, combining a multi-population evolutionary algorithm with a QAOA. We propose to distribute independent population instances on two separate Quantum Processing Units (QPUs) orchestrated by a scheduler. This strategy is novel to the best of the authors' knowledge: numerous scheduling prescriptions have been proposed by the community~\cite{Perlin, Doi, Doi1}; however, none are tailored to parallelizing MGAs across QPUs.
  
This paper reports on the solution accuracy and variance of our EA compared to the traditional Constrained Optimization BY Linear Approximation (COBYLA) method \cite{Powell1994, 1998AcNum...7..287P, powell2007view} used with the QAOA. Tests were performed on regular $d$-3 graphs both on the simulator and real quantum hardware. The paper is structured as follows. The methodology in Sec.~\ref{sec:method} provides a focused overview of the QAOA for a Max-Cut problem and the relevant elements for the EA. Sec.~\ref{sec:setup} outlines the computational facilities used and the parameters chosen in our experiments. Next, we show the results in Sec.~\ref{sec:results}, together with their discussion. Lastly, Sec.~\ref{sec:conclusion} summarizes the work presented and its limitations and concludes with an outlook on future work.

\section{Methodology}
\label{sec:method}

In this study, we propose a new hybrid framework, termed the \textit{Evolutionary Quantum Approximate Optimization Algorithm} (E-QAOA), which combines variational quantum algorithms and classical evolutionary optimization. The methodology is organized into three main components, described as follows.
\begin{enumerate}
    \item Formulation of the Max-Cut problem, a well-known combinatorial optimization task.
    \item Design of the QAOA circuit, consisting of state preparation, unitary evolution, and post-processing of measurement outcomes.
    \item Evolutionary strategies used to optimize the variational parameters, discussing key concepts and terminologies from evolutionary and genetic algorithms.
\end{enumerate}

Notably, in the evolutionary computation literature, the terms \textit{cost} and \textit{fitness} are used interchangeably; both refer to the objective function guiding the optimization.

\subsection{Max-Cut}
\label{sec:method:max-cut}

As demonstrated in the seminal work by Farhi et al.~\cite{Farhi}, QAOA can be adapted to solve the Max-Cut problem by optimizing a variational quantum circuit. For an unweighted graph $G=(V,E)$ with $n$ nodes, the goal is to partition $V$ into two subsets that maximize the number of edges between them. Equivalently, the objective is to maximize $C(x) = 1/2\,\sum_{(i,j) \in E} 1 - (1-2x_i)(1-2x_j)$,
where each $x_i \in \{0,1\}$ indicates the partition assignment of node $i$. In the quantum setting, nodes are mapped one-to-one to qubits, and the algorithm samples bit-strings $x \in \{0,1\}^n$, with each bit representing the corresponding node's partition.

\subsection{The QAOA circuit}
\label{sec:method:circuit}

\begin{figure*}[ht!]
    \centering
    \includegraphics[width=\linewidth]{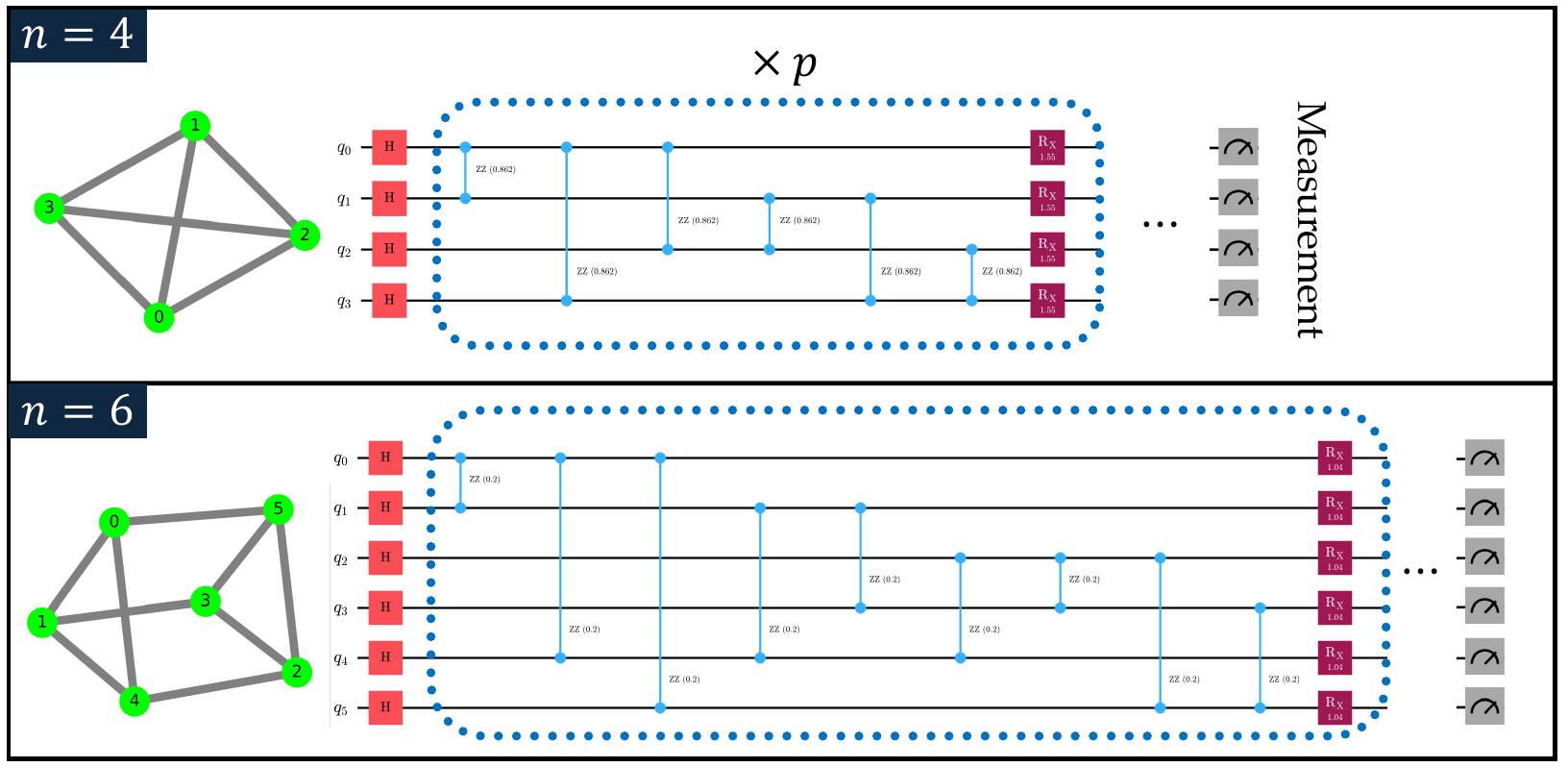}
    
    \caption{Example circuits for $d$-3 regular graphs with 4 (top) and 6 (bottom) nodes. The circuits have as many qubits as there are in the respective graphs, and are structured as follows: a Hadamard gate ($H$, in red) is first applied to all qubits, followed by $RZZ$ gates constituting the connectivity section and $Rx$ gates on all qubits. This circuit section (dotted-line rectangle) is repeated $p$ times before the qubit states are measured. Crucially, increasing either $p$ or the number of edges, i.e. the graph connectivity, increases the circuit depth.}
    \label{fig:diagrams}
\end{figure*}

We adopt the standard QAOA, as implemented in Qiskit~\cite{javadi2024quantum}, to analyze the performance improvement when combined with an EA. In our implementation, the circuit alternates between two parameterized unitary operators. The mixer and cost unitaries are given by
\begin{align}
    U(H_M) &= e^{-i\beta H_M} = e^{-i\beta X_0}e^{-i\beta X_1} \cdots e^{-i\beta X_n} \\
    U(H_C) &= e^{-i\gamma H_C} = e^{-i\gamma Z_i Z_j}e^{-i\gamma Z_j Z_k} \cdots e^{-i\gamma Z_k Z_i},
\end{align}
respectively, where $X_i$ and $Z_i$ denote the Pauli-$X$ and -$Z$ operators on the $i^{\rm th}$ qubit, and the corresponding $RZZ$ gates are applied to each edge $(i,j)$ in the graph. These layers are repeated $p$ times (with each layer having distinct parameters $\beta$ and $\gamma$, initialized randomly on $(-\pi,\pi]$). A schematic of the circuit for a given graph is presented in Fig.~\ref{fig:diagrams}. In the limit $p\to\infty$, QAOA converges to the optimal solution and finds the global cost minimum under ideal parameter optimization. 

Finally, we use COBYLA (via the \texttt{minimize} method in \textsc{Scipy}~\cite{virtanen2020scipy}) to update $\beta$ and $\gamma$ as our control experiment. COBYLA has been shown to provide generally good convergence, on par with other common optimizers like SLSQP, L‑BFGS‑B \cite{10.1145/3583133.3596396}. For this reason, we restrict our investigation to COBYLA, with a plan to compare EAs to other optimizer choices (e.g., as in \cite{9951304}) in future work.

\subsection{Fitness evaluation}
\label{sec:method:fit-eval}

In our experiments, we employ two metrics to evaluate the fitness of candidate solutions and update the QAOA parameters $\beta$ and $\gamma$. First, fitness is defined as the Max-Cut value corresponding to the most frequent bit-string from the measurement distribution, obtained via Qiskit's \texttt{max\_count} feature. Second, we use Conditional Value-at-Risk (CVaR)~\cite{Barkoutsos}, which computes the expectation value over the best $\alpha$ fraction of outcomes (with $\alpha\in(0,1]$, where $\alpha=1$ yields the standard expectation). Empirical studies have shown that CVaR leads to faster convergence on both simulators and quantum hardware~\cite{Kolotouros} with respect to using the standard expectation value, as the optimization is driven by the expectation of the $\alpha$-tail of the distribution rather than the whole distribution, or a single bit-string (\texttt{max\_count}). In the latter method, the number of shots required to reliably converge using the most frequent occurring bit-string would scale combinatorially with the number of nodes for our problem, quickly saturating the capabilities of near-term quantum hardware and ultimately returning a randomly selected bit-string due to noise. For this reason, we only show results using the \texttt{max\_count} bit-string selection in Fig.~\ref{fig:cvar-vs-not} to illustrate the comparison with CVaR, and use the CVaR method throughout the rest of the paper.

To compare performance across various graph sizes, we define the approximation ratio, $\mathcal{R}$, as
\begin{equation}
    \mathcal{R} = \frac{C({\rm solution~found})}{C({\rm optimal~solution})},
\label{eq:approx_ratio}
\end{equation}
which equals 1 when the optimal solution is achieved.

\subsection{Evolutionary algorithm}
\label{sec:method:evolutionary}
Our EA implementation features the core rules of evolutionary dynamics: parent selection, recombination operator, mutation, and survivor selection. These components ensure a balance between exploration of the cost (or fitness) landscape and exploitation around candidate solutions. Additional heuristic rules and properties can be added to the EA to improve the convergence performance (e.g. replacement and termination strategy), however, these are less crucial than the core rules and will not be investigated in this work.

The process begins with a population of $n_{\rm pop}$ candidate solutions, denoted by $Y = \{ Y_1, Y_2,\cdots, Y_{n_{\rm pop}} \}$, each encoded as a `genotype' (or chromosome) consisting of real-valued parameters:
\begin{equation}
    Y_i = [\beta_1,\gamma_1, \beta_2,\gamma_2,\ldots,\beta_p,\gamma_p],
\label{eq:chromosome}
\end{equation}
with each gene (or allele) initialized randomly in $[-\pi, \pi]$. 

Given these initial conditions, the algorithm updates the population by selecting `parent' solutions to produce `offsprings'. Parents are selected using Stochastic Universal Sampling (SUS)~\cite{Baker}, where the selection probability is proportional to the normalized fitness values. This ensures that all individuals $Y_i$ have a chance to maintain their genes through the following generation, except for the least fit solution. We keep the population size constant by imposing that two parents generate two new offspring. SUS allows sampling with replacement, where the fittest individual can be chosen in multiple (or all) pairs of parents but can only be selected once for each pair of parents, and thus they cannot pair with themselves. Once the parents are chosen, the offspring are produced via recombination and mutation.

Recombination was applied $100\%$ of the time on each pair of parents through whole arithmetic crossover, i.e. on all of their genes. This operation is described as follows. Given two parents, $Y_i$ and $Y_j$, one of their offspring $Y_{i}^{\prime}$ is given by
\begin{equation}
\begin{split}
     Y^\prime_{i} = [~&\alpha\beta_{Y_{i},1} + (1-\alpha)\beta_{Y_{j},1},  \alpha\gamma_{Y_{i},1} + (1-\alpha)\gamma_{Y_{j},1},  \\
     &\cdots, \\&\alpha\beta_{Y_{i},p} + (1-\alpha)\beta_{Y_{j},p},  \alpha\gamma_{Y_{i},p} + (1-\alpha)\gamma_{Y_{j},p}~],
\end{split}
\end{equation}
where $\alpha$ is a randomly generated variable uniform on $[0,1]$ (not to be confused with the CVaR $\alpha$ parameter). 

Mutation is then applied to each of the genes with a probability $p_{\sigma}=0.2$. This heuristic strategy is, once again, inspired by Darwinian natural evolution, where offspring are most likely never the exact combination of their parents but are subject to random mutation on their genome. In our EA, we use self-adaptive mutation, where we add $2p$ different $\sigma$ mutation values within the genotype, such that they also undergo recombination and mutation through the generations, and they evolve alongside the weights $\beta$ and $\gamma$. By attaching these $\sigma$ values, the genotypes become
\begin{equation}
    Y_i = [\beta_1, \gamma_1, \beta_2, \gamma_2, \cdots, \beta_p, \gamma_p, ~ \sigma_1, \sigma_2,\cdots,\sigma_{2p}] \, .
\label{eq:chromosome_mutation}
\end{equation}
Each $\sigma$ is normally distributed on $[0,1]$ and mutation occurs as follows:
\begin{gather}
    \sigma_i^\prime = \sigma_{i}\, e^{\tau^{\prime}\, N(0,1)+ \tau\,  N_i(0,1)} \\
    \downarrow \notag \\
    \beta_i^\prime = \beta_i + \sigma_{i}^{\prime}\, N(0,1) \\
    \gamma_{i}^{\prime} = \gamma_i + \sigma_{i+1}^{\prime}\, N(0,1) \, .
\label{eq:mutation}
\end{gather}

The variables $\tau$ and $\tau^\prime$ are usually set to be inversely proportional to the square and fourth root of the population size, respectively, but can be defined by the user, and can be interpreted as a learning rate~\cite{Eiben}. Furthermore, $\sigma$ is mutated by a log-normal distribution to meet certain requirements, like smaller modifications, the equal likelihood of drawing a value and its reciprocal~\cite{Bäck}. After these values evolve we enforce a minimum threshold such that $|\sigma^{\prime}| \geq \sigma_{\rm min}$.  If recombination or mutation parameters push $\beta$ and $\gamma$ out of the predefined ranges, we enforce periodic boundaries to return the genes to their proper domain. 

After creating the new generation, we implement generational survivor selection, where $\mu$ fittest individuals from the older generation will be preserved in the newer generation if no offspring can surpass their fitness values. This ensures that the best-performing individuals are not lost in evolution and that their genes continue to be passed on through generations. This strategy is particularly useful in the presence of noise, like currently available quantum hardware: even if the population's average fitness decreases, the next generations can retain and spread the fittest genes again. The operations described are repeated in this order for a fixed number of generations, $g$. Finally, the pseudo-code for our E-QAOA is summarized in Algorithm~\ref{alg:evo_qaoa}.

\begin{algorithm}
\caption{Steps to implement this work's EA for evolving a population of solutions.}
\label{alg:evo_qaoa}
\begin{algorithmic}
\State $Y \gets  [Y_1, Y_2, \cdots, Y_{n_{\rm pop}}]$
\State $g \gets 0$
\While{$\texttt{generations} < g$}
    \For{\texttt{i in (1, $n_{\rm pop}$)}}
        \State $\texttt{fitness$_{Y_i}$} \gets \texttt{f(QAOA($Y_i$))}$
        \State $\texttt{parents} \gets \texttt{SUS(Y,fitness$_Y$)}$
        \State $Y_{i}^{'} \gets \texttt{recombination(parent$_i$, parent$_j$)}$ 
        \For{\texttt{$j$ in range(2$p$)}}
            \State $Y_{i}^{\prime}[j] \gets \texttt{mutation($Y_{i}^{\prime}[j], p_{\sigma}$)}$
        \EndFor
    \EndFor
    \State $Y' \gets \texttt{elitism(parents, $Y^\prime$, $\mu$)}$
    \State $g \gets g+1$
\EndWhile
\end{algorithmic}
\end{algorithm}

The approach for a multi-population EA is similar. In our set-up, we consider multiple QPUs, each evolving an independent population of solutions, as described above, for a fraction $g_{\rm f}$ of the total number of generations, e.g. $g_{\rm f}=g/3$. At every $g_{\rm f}$ iterations, the fittest individuals from each population \textit{migrate} to their non-native population. In our computational set-up, the fittest individuals are copied to the target device to replace the weakest individuals and maintain a constant population size. At migration time, the fittest individuals to migrate are selected based on the absolute best Max-Cut value in the distribution rather than their $\alpha$-tail fitness given by CVaR. The weakest individuals are also evaluated this way before being replaced. Then, the populations continue their evolution independently until the next migration or upon reaching the maximum number of iterations required for termination. 

When implemented on a real quantum-classical infrastructure, migration occurs classically so that the job is re-queued. Therefore, the overall execution time is impacted by the overall system usage, network traffic and, where necessary, manual scheduling of restart jobs. 

As mentioned above, one of the benefits of an MGA approach is an increased diversity within a population of solutions, boosting the chances of guiding the algorithm towards optimality. We measure diversity by quantifying the uniqueness of values, in particular the number of unique fitness values and unique genes across the population. We normalize this value by the size of the population as follows:

\begin{equation}
    \textrm{Uniqueness~ratio} = \frac{|\{\beta(Y_i), \gamma(Y_i): Y_i\in Y\}|}{n_{\rm pop}},
\label{eq:uniqueness}
\end{equation}
where $|\cdot|$ denotes the umber of unique gene values found in the population.

\section{Set-up}
\label{sec:setup}

The simulations are conducted on the \textit{Scafell Pike} high-performance computing system at Sci-Tech Daresbury (UK), which features dual-socket compute units with 16-core Intel Xeon Gold CPUs per socket. Hardware experiments are executed on IBM's 127-qubit Eagle processors \textit{ibm\_cusco} and \textit{ibm\_nazca}, which exhibit an Error-Per-Layered-Gate (EPLG) on a 100-qubit chain of 5.9\% and 3.2\%, respectively at the time when the calculations are executed~\cite{2023arXiv231105933M}. For each fitness evaluation, the QAOA circuit is sampled with \(10^4\) shots. We tested experiments (not shown here) varying the number of shots and found that \(10^4\) provides a suitable convergence for our problem size with the quantum hardware available. For reproducibility, all experimental parameters are summarized in Table~\ref{tab:EA} and remain unchanged across experiments.

\begin{table}[bp]
\centering
\caption{Summary of variables used in our E-QAOA implementation, their description, and values.}
\label{tab:EA}
\begin{tabular}{clc}
    \toprule
    \textbf{Variables} &  \textbf{Description} & \textbf{Value} \vspace{3pt}\\
    \toprule
    $n$ & Number of nodes (graph size) & $[4, \dots ,26]$ \\
    $n_{\rm pop}$ & Number of individuals in a population & $[8, \dots ,20]$ \\
    $g$ & Number of generations for termination & \{10, 20\} \\
    $g_{\rm f}$ & Number of generations for migration & \{5, 7\} \\
    $p$ & Number of rounds for the QAOA circuit & 2\\
    $\gamma$, $\beta$ & Ansatz parameters to optimize& $[-\pi$, $\pi]$ \\
    $\alpha$ & CVaR value & 0.15 \\
    $p_{\sigma}$ & Probability of mutation & 0.2 \\
    $\sigma$ & Mutation parameter & $N(0,1)$ \\
    $\sigma_{\rm min}$ & Minimum absolute value for $\sigma$ & 0.1 \\
    $\mu$ & Number of elite & 1 \\
    $\tau$ & Parameter used for mutation & $\sqrt{2}/ 2 \cdot {n_{\rm pop}}^{-1/4}$ \\ 
    $\tau^{\prime}$ & Parameter used for mutation & $\sqrt{2}/ 2 \cdot {n_{\rm pop}}^{-1/2}$ \vspace{3pt}\\ 
    \bottomrule
\end{tabular}
\end{table}

\section{Results and Discussion}
\label{sec:results}
Now, we present the results of our experiments and a discussion on their performance.
Tests are first run on a classical state vector simulator to evaluate potential advantages and tune the appropriate parameters, ahead of the hardware experiments. Then, they are run on \textit{ibm\_cusco} and \textit{ibm\_nazca}, the IBM Quantum devices available to the authors via the cloud.

\subsection{Simulations}

Fig.~\ref{fig:cvar-vs-not} shows the performance of our EA optimizer compared to COBYLA's for $d$-3 regular graphs sized between 4 and 26 nodes. To ensure fair comparison between the methods, we compute the average and standard deviation of 10 independent COBYLA-based runs with E-QAOA runs with a population size $n_{pop}=10$. In the top panel, we note that both methods achieve the optimal solution, i.e. an approximation ratio of 1 (dotted horizontal line), for small graphs, as expected. However, from $n \geq 16$, both optimizers show a decline in accuracy: the EA maintains an advantage over COBYLA for accuracy and variance. Fig.~\ref{fig:cvar-vs-not} clearly shows that the CVaR method is more stable than \texttt{max\_count}, as expected from prior considerations. The results with \texttt{max\_count} are shown in the bottom panel, where the approximation ratio reaches 1 fewer times than for the CVaR case. The $n=12$ graph is an interesting case, where the \texttt{max\_count} only achieves an approximation ratio of $\approx0.5$, compatible with random guessing. The variance for COBYLA is substantially larger than for E-QAOA, and the mean accuracy appears lower than our method ( $\approx\, 0.61$ for both $n=20$ and $26$). Conversely, the EA optimizer appears more robust, even though the accuracy does not remain as high as with CVaR, the variance increases only slightly, and it is still able to maintain an approximation ratio at  $\approx\, 0.77$ or above for all cases, with a much lower discrepancy between repetitions than COBYLA. Note that the function \texttt{max\_count} is equivalent to using the CVaR function with $\alpha = 1 / \text{shots}$, such that reducing $\alpha$ drastically will create instability in convergence. Through experimentation we found the value of $\alpha=0.15$ to be robust enough for our experiments.

\begin{figure}
    \centering
    \includegraphics[width=\linewidth]{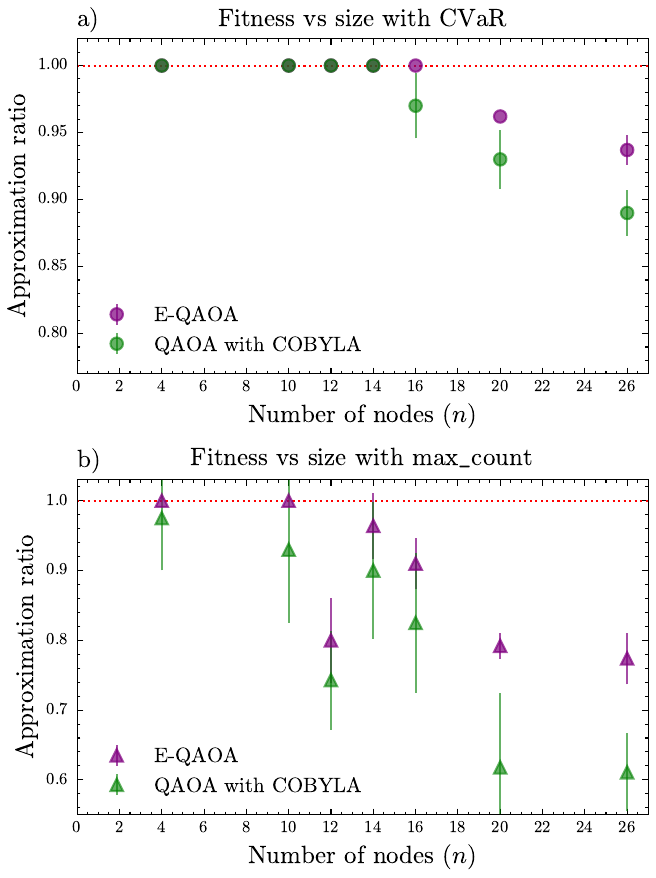}
    
    \caption{Approximation ratio (Eq.~\ref{eq:approx_ratio}) averaged over 10 runs per graph, for graph sizes $n=[4,10,12,14,16,20,26]$ on $d$-3 regular graphs solved using QAOA paired with either COBYLA (green) or the EA (purple). We show results using CVaR (top) and \texttt{max\_count} (bottom) for the fitness of solutions. In the EA, we set $g=10$ and $n_{\rm pop}=10$ and the COBYLA-based runs are computed from 10 independent runs run for 10 iterations. The error bars indicate the standard deviation of 10 realizations. The horizontal dotted line represents the maximum approximation ratio achievable.}
    \label{fig:cvar-vs-not}
\end{figure}

We expect that the higher robustness for the EA is due to an evolving population of solutions rather than just an individual realization; additionally, our runs appear to suggest that the EA is less susceptible to initial conditions, e.g., due to a poor initial guess. Furthermore, the EA can explore a wider search space within the fitness landscape instead of being guided by gradients and is thus able to consistently find more accurate solutions.  

\begin{figure*}[ht]
    \centering
    \includegraphics[width=\linewidth]{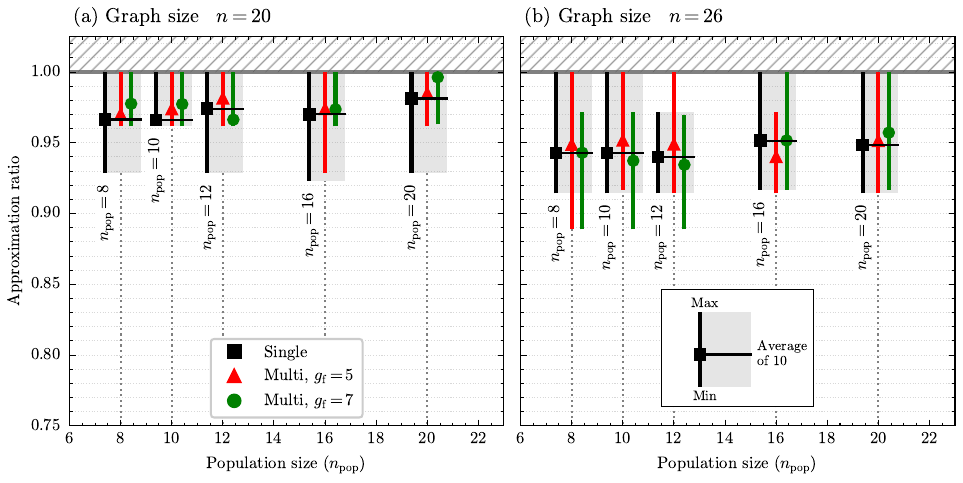}
    
    \caption{Simulated approximation ratio as a function of the population size ($n_{\rm pop}$) for 20-node (left) and 26-node (right) graphs. Classical simulations were run for values of $n_{\rm pop}=[8,10,12,16,20]$, indicated by the vertical dotted lines with labels. For a given $n_{\rm pop}$, the approximation ratio for single-population runs (black) is compared to multi-population-EA results for $g_{\rm f}=5$ (red) and $g_{\rm f}=7$ (green). Markers represent the average approximation ratio from 10 distinct independent regular $d$-3 random graphs, and the error bars span the minimum and maximum obtained for each $n_{\rm pop}$. The black and green markers are slightly shifted along the $x$-axis to facilitate the comparison; as guidelines, we show solid horizontal lines (and grey boxes) next to the black markers to compare the mean multi-population ratios (and the extent) with the single-population results. The hatched area for ratios above 1 is excluded.}
    \label{fig:sim_multipop}
\end{figure*}

In Fig.~\ref{fig:sim_multipop}, we show the simulations for finding the average Max-Cut approximation ratio on a graph with $n=20$ nodes on the left, and $n=26$ on the right panel for a single-population algorithm versus a multi-population one as explained in the methodology. Note that the population size is the same for each isolated population; therefore, the MGA approach is effectively evolving double the number of individuals independently. In the left panel, the multi-population algorithm for $g_{\rm f}=5$, shown in red, achieves a slightly better solution quality than the single population for all combinations of population size, over an average of 10 random graphs. For both the single and multi-strategy, the mean approximation ratio tends to increase as the population size increases, except for $n_{\rm pop}=16$. This is expected as evaluating more individuals by increasing the population size should enhance convergence, albeit not necessarily in a linear fashion. For $g_{\rm f}=7$, this advantage slightly increases over the one achieved with $g_{\rm f}=5$, in four out of the 5 instances. Beyond achieving a better mean solution, the multi-population strategies also achieve a higher minimum for all experiments. Another interesting thing to observe is that the multi-population algorithm, for both $g_{\rm f}$ values, will, in most cases shown, achieve at least the same accuracy rate as the single-population algorithm with a larger population size. This is except for $n_{\rm pop}=16$ again, yet, in the case of $n_{\rm pop}=12$ the multi-population approach with $g_{\rm f}=5$ achieves an accuracy better than the single population one with $n_{\rm pop}=20$. This result suggests there is potential for a speed-up for the MGA approach, but such an investigation of that performance is outside the scope of this paper.  
  
On the other hand, the right panel illustrating results for $n=26$ suggests that further tuning is required between $n_{\rm pop}$ and $g_{\rm f}$ to achieve better approximation ratios. For the multi-population with $g_{\rm f}=5$, we still achieve better or on-par results, with a lower population size than the single population does with $n_{\rm pop}=20$. A larger $g_{\rm f}$ value only helps with larger population sizes, otherwise, it is detrimental. In these cases, for $n_{\rm pop} \leq 12$, waiting for a larger number of generations might hinder successful evolution and convergence as we introduce a new, fresh individual when the algorithm has already started converging towards another set of genes. This competition between two very fit individuals might split a small population in a way that does not explore the fitness landscape optimally. In both plots, the multi-population algorithm with $g_{\rm f}=7$ and $n_{\rm pop}=20$ achieves the highest average approximation ratio overall.  

\subsection{Hardware}
Following promising simulation results, we validate our approach experimentally on the \textit{ibm\_nazca} quantum device by solving the Max-Cut problem for 5 random $d$-3 regular graphs of size $n=16$. As in the simulation, the optimal Max-Cut solution is found in all cases. Subsequently, we scale to graphs with $n=20$ vertices and a population size of $n_{\rm pop}=12$. This system is selected because it produced the highest approximation ratio and quickest convergence for classical state vector simulations multi-population strategy. We show results for a single and multi-population approach with $g_{\rm f}=5$ in Fig.~\ref{fig:hardware}. The multi-population strategy achieves a slightly higher mean and a higher maximum on hardware than the single population. Both methods perform slightly worse on hardware than in simulation. However, this effect is expected, and the small discrepancy between hardware and simulations suggests that (i) the method could be successfully run on larger systems, and (ii) the improvement of MGA over single population is maintained on hardware. Due to constraints on available QPU time, we simulate these systems on the hardware platform for 15 generations rather than 20, and test 5 out of 10 random graphs. While this restriction may mitigate claims regarding the advantages of E-QAOA and multi-population strategies, we present the hardware results as useful proof-of-concept research and a robust validation of our classical simulations. 

\begin{figure}
    \centering
    \includegraphics[width=\linewidth]{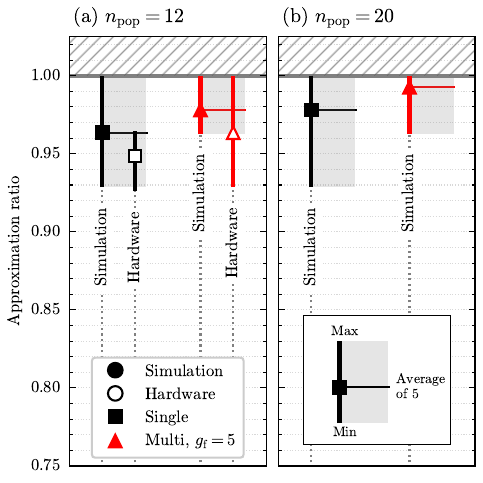}
    
    \caption{As in Fig.~\ref{fig:sim_multipop}, but for hardware runs (empty markers) of 5 random regular $d$-3 graphs with $n=20$. Hardware runs configured with $n_{\rm pop}=12$, $g=15$ and $g_{\rm f}=5$. The single-population run (empty square) was performed on \textit{ibm\_nazca} and the multi-population run (empty triangle) was performed on both \textit{ibm\_nazca} and \textit{ibm\_cusco}. For comparison, we show simulation results for $n_{\rm pop}=12$ and 20 as filled markers. The average is computed over 5 different graphs.}
    \label{fig:hardware}
\end{figure}

\begin{figure*}
    \centering
    \includegraphics[width=\linewidth]{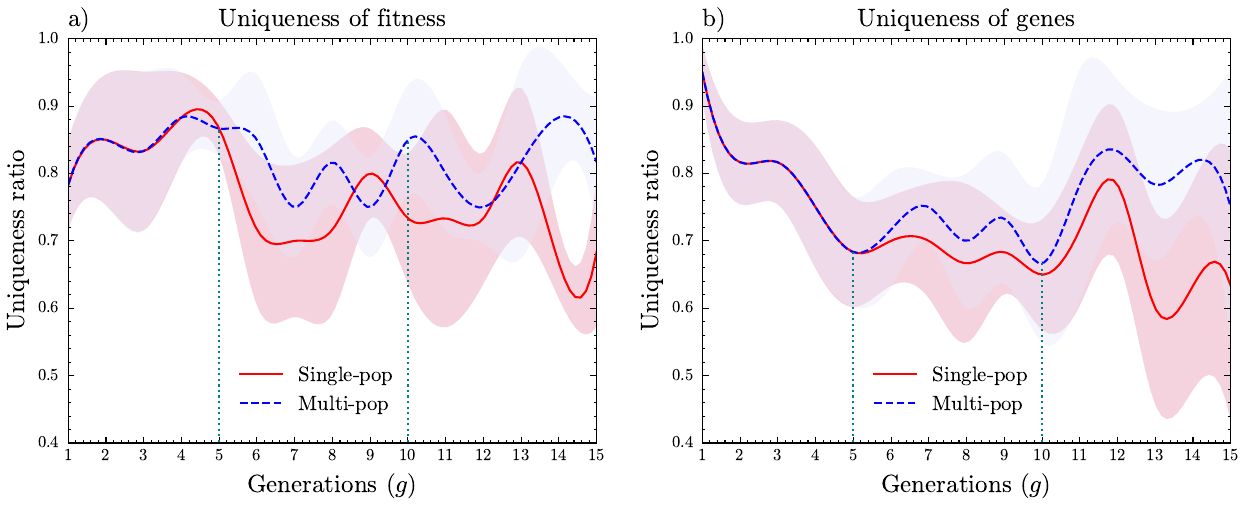}
    
    \caption{Both figures show a uniqueness ratio over the evolution of generations for a single-population (solid red) and a multi-population (dashed blue) on the average of 5 random graphs of size $n=20$. For all hardware runs $g=15$, $g_{\rm f}=5$ and $n_{\rm pop}=12$. These statistics are achieved on the \textit{ibm\_nazca} machine, with the shaded area representing the standard deviation. Sub-figure a) shows the uniqueness in fitness values calculated with CVaR. Sub-figure b) shows the uniqueness in the first allele of the genetic code, i.e. $\beta_0$. Note that for the multi-population approach, the uniqueness is calculated only on one population, within the \textit{ibm\_nazca} machine; as such, the values between the two approaches overlap before the first migration. The vertical dotted lines represent the generation number at which these migrations happen. Finally, we use a spline function to smooth the uniqueness ratio values between (discrete) generations.}
    \label{fig:uniqueness}
\end{figure*}

A multi-population approach can enhance the diversity within the genes of a population, avoid premature convergence, and search the fitness landscape for values that might not otherwise be reached with a single population. Also note that the heterogeneity of different devices does not have a negative impact on the evolution of populations, rather it can be used to promote healthy diversity that aids more effective searching of the solution space. Fig.~\ref{fig:uniqueness} shows how the uniqueness for two different parameters evolves for the single and multi-population approaches as tested on hardware. The left panel (a) shows the uniqueness with respect to how many fitness values are found within the \textit{ibm\_nazca} population. Here, the multi-population approach trends slightly higher whilst intercrossing with the single-population approach. Given that the fitness value is not related in a one-to-one fashion to the genes or ansatz parameters of the circuit, the fitness value cannot show the whole picture on its own, since a family of $\beta$ and $\gamma$ combinations would produce the same CVaR output when evaluating the fitness. To provide more complete information, we also show the uniqueness of $\beta_0$ values in the right panel (b). In this case, the separation is more pronounced, with the migrations, shown in the vertical dotted lines happening at regular intervals, pushing and maintaining a higher uniqueness ratio within the multi-population approach. The comparison is performed using only one allele as the cross-over operator for recombination, which occurs on all genes by construction, is expected to generate unique individuals in these terms without mutation. As for the uniqueness of fitness values, neither does the diversity between genes provide the whole picture, requiring further analysis of the expressivity of the circuit for the genes, and a fitness value using only the best Max-Cut value within the distribution, i.e. $\alpha = 10^{-4}$, or 1 count in $10^4$ shots.  
  
Given these results, we can conclude that the multi-population E-QAOA has potential but may need careful parameter tuning to produce optimal results and successfully compete with the single-population E-QAOA strategy. In this respect, we would allow further generations to evolve and converge without migrations once successful swaps have already modified the distributed populations. This further step is necessary to ultimately allow diversity within the population to fall and let individual solutions converge towards the best genetic codes.

\section{Conclusion}
\label{sec:conclusion}

This work introduces a multi-population evolutionary QAOA framework for the Max-Cut problem, integrating an evolutionary algorithm into QAOA's variational parameter optimization protocol. In Fig.~\ref{fig:cvar-vs-not}, we benchmark this evolutionary QAOA approach against a standard COBYLA-based QAOA on random regular graphs (up to 26 nodes) in classical simulation, and in Figs.~\ref{fig:hardware} and~\ref{fig:sim_multipop}, we run similar calculations on real IBM quantum devices. 

The approach consistently attained equal or higher approximation ratios than COBYLA, for a given number of iterations, while yielding significantly lower cost variance, indicating reliable performance and reproducibility. Notably, employing the CVaR (Conditional Value at Risk) metric as the objective expectation from the tail of the sampled distribution enhanced outcome stability and robustness to noise, resulting in more consistent solution accuracy. In Fig.~\ref{fig:sim_multipop}, we demonstrate a distributed execution of the multi-population E-QAOA by running two isolated populations in parallel on separate QPUs with periodic migration of elite solutions between them every $g_{\rm f}=5$ generations. This strategy preserved greater genetic diversity and outperformed the single-population baseline (Fig.~\ref{fig:uniqueness}), confirming the viability of parallel population evolution on actual quantum processors. The hardware results align with simulations (aside from expected noise-induced degradation), showing that our method's advantages carry over to real quantum devices. Overall, our study demonstrates a distributed QAOA paradigm that improves (i) solution accuracy, (ii) reliability over conventional single-optimizer approaches, and (iii) can scale on heterogeneous networks of quantum processors. 

These findings lay a foundation for future work on parallel and distributed quantum optimization, such as scaling to larger problem instances beyond 100 qubits on multiple QPUs, refining inter-population communication schemes (see also \cite{carrera2024combining, andres2024distributing, main2025distributed}). Investigations into adaptive evolutionary strategies, such as dynamic migration intervals, topology-aware elite exchanges, and hybrid evolutionary-gradient optimization methods, could further accelerate convergence rates and maintain genetic diversity. Additionally, extending the framework to other combinatorial optimization tasks beyond Max-Cut, such as portfolio optimization, scheduling, and constrained optimization, is a natural next step, given QAOA's versatility in addressing various QUBO-formulated problems. Finally, results from different quantum hardware technologies and providers should be compared and benchmarked. We envisage significantly more accurate and better-converged results in the upcoming early fault-tolerant regime, where hardware topologies can handle densely connected graphs and sample distributions at a higher frequency ($\approx200$ kHz for IBM Heron devices).

Finally, proposed distributed E-QAOA is particularly suited for graph-based optimization problems in domains like network design \cite{lin2009quantum}, quantum machine learning (feature selection, hyperparameter tuning, or quantum data encoding) \cite{rawat2024automated, 2025MLS&T...6a5056M}, quantum chemistry (molecular electronic structure and ground-state energy prediction) \cite{supady2015first, boy2024energy}, and high-energy physics \cite{di2024quantum, 2025MLS&T...6a5056M}. The combined benefits of parallel search and evolutionary robustness position E-QAOA as a valuable tool in these demanding quantum optimization applications. 

\section*{Acknowledgment}
This work was supported by the Hartree National Centre for Digital Innovation, a UK Government-funded collaboration between STFC and IBM. IBM, the IBM logo, and \href{https://www.ibm.com}{www.ibm.com} are trademarks of International Business Machines Corp., registered in many jurisdictions worldwide. Other product and service names might be trademarks of IBM or other companies. The current list of IBM trademarks is available at \href{https://www.ibm.com/legal/copytrade}{www.ibm.com/legal/copytrade}.
The research in this paper made use of the following software packages and libraries: 
\textsc{Python}~\cite{van1995python},
\textsc{Qiskit}~\cite{javadi2024quantum},
\textsc{Numpy}~\cite{harris2020array},
\textsc{Scipy}~\cite{virtanen2020scipy},
\textsc{Random}~\cite{random},
\textsc{NetworkX}~\cite{osti_960616, hagberg2020networkx},
\textsc{Matplotlib}~\cite{hunter2007matplotlib, caswell2020matplotlib}.

\bibliographystyle{IEEEtran}
\bibliography{main}

\end{document}